# Unidirectional Alignment of AgCN Microwires on Distorted Transition Metal Dichalcogenide Crystals

Myeongjin Jang[1,2], Hyeonhu Bae[3], Yangjin Lee[1,2], Woongki Na[4], Byungkyu Yu[2], Soyeon Choi[1], Hyeonsik Cheong[4], Hoonkyung Lee[3], and Kwanpyo Kim[1,2,*]

[1]Department of Physics, Yonsei University, Seoul 03722, Korea

[2]Center for Nanomedicine, Institute for Basic Science (IBS), Seoul 03722, Korea

[3]Department of Physics, Konkuk University, Seoul 05029, Korea

[4]Department of Physics, Sogang University, Seoul 04107, Korea

[*]Address correspondence to K.K. (kpkim@yonsei.ac.kr)

**Abstract:** Van der Waals epitaxy on the surface of two-dimensional (2D) layered crystals has gained significant research interest for the assembly of well-ordered nanostructures and fabrication of vertical heterostructures based on 2D crystals. Although van der Waals epitaxial assembly on the hexagonal phase of transition metal dichalcogenides (TMDCs) has been relatively well characterized, a comparable study on the distorted octahedral phase (1T' or $T_d$) of TMDCs is largely lacking. Here we investigate the assembly behavior of one-dimensional (1D) AgCN microwires on various distorted TMDC crystals, namely 1T'-$MoTe_2$, $T_d$-$WTe_2$, and 1T'-$ReS_2$. The unidirectional alignment of AgCN chains is observed on these crystals, reflecting the symmetry of underlying distorted TMDCs. Polarized Raman spectroscopy and transmission electron microscopy directly confirm that AgCN chains display the remarkable alignment behavior along the distorted chain directions of underlying TMDCs. The observed unidirectional assembly behavior can be attributed to the favorable adsorption configurations of 1D chains along the substrate distortion, which is supported by our theoretical calculations and observation of similar assembly behavior from different cyanide chains. The aligned AgCN microwires can be harnessed as facile markers to identify polymorphs and crystal orientations of TMDCs.

**INTRODUCTION**

Transition metal dichalcogenides (TMDCs) have recently gained widespread research interest for their novel properties[1-4] as well as for applications in various electrical and chemical processes[5-6]. Depending on its species of transition metals and chalcogenides, TMDCs can assume various crystal structures, including 2H, 1T, 1T', and $T_d$ phases[5, 7-9]. Structural analysis of these materials along with their influence on the crystal's electrical and chemical properties have been important subjects[5]. For example, certain TMDCs, such as $MoTe_2$, exhibit multiple accessible polymorph structures, and their structural phase transitions controlled by pressure, temperature, and doping have been utilized to induce the semiconductor-metal transition and to achieve better electrical contact properties[10-11]. Among various TMDC crystal structures, the family of distorted octahedral (1T' or $T_d$) crystals has recently emerged as exciting platforms to investigate various physical phenomena[12-13]. For instance, $T_d$ phase of and $WTe_2$ and $MoTe_2$ have become important playgrounds for hosting Weyl fermions[14-18], and 1T'-$ReS_2$ is known to exhibit monolayer-like properties due to small interlayer coupling even in bulk samples[12].

Owing to the distorted crystal structure, 1T' (or $T_d$) TMDCs possess strong in-plane physical and electronic anisotropy[19-25]. These distorted TMDC crystals exhibit the puckered surface structure along one direction and the anisotropic surface topography may induce unique assembly behaviors of nanomaterials. Previous studies have examined the potential of various hexagonal 2D crystals, including graphene, h-BN, and 2H-TMDCs, as templates for van der Waals assembly of various nanostructures[26-33]. The assembled nanostructures and resulting heterostructures have shown various synergetic electrical and optical properties arising from strong interactions between the substrate and the assembled structures[32-35]. In particular, various types of assembly phenomena, including one-dimensional (1D) assembly on hexagonal 2D crystals, have previously been reported[28, 35-36]. On the other hand, the observation of comparable phenomena on distorted TMDCs (1T' or $T_d$ phase) is largely lacking. In particular, the assembly behavior of 1D chains on distorted TMDCs is yet to be reported.

Here we investigate the assembly behavior of AgCN microwires on various distorted TMDC crystals (1T'-MoTe$_2$, $T_d$-WTe$_2$, and 1T'-ReS$_2$). The AgCN crystal is a close-packed structure of atomic wires (Ag-C≡N)n as shown in Figure 1a[36-37]. AgCN can interact with 2D substrates mainly via weak van der Waals interaction and serve as an ideal 1D component for assembly (Figure 1b). We observed the unidirectional alignment of AgCN microwires on distorted TMDC crystals. Polarized Raman spectroscopy and transmission electron microscopy (TEM) confirmed the crystallographic alignment between the AgCN microwire axis and the distorted lattice direction of underlying 1T' or $T_d$ crystals. We also performed theoretical calculations on the adsorption energy of AgCN chains on a distorted 1T' crystal to understand the mechanisms of the observed unidirectional alignment. Our work demonstrates that distortion-mediated van der Waals epitaxy can be a useful way to induce anisotropic but highly ordered nanostructures.

**RESULTS AND DISCUSSION**

To analyze the effect of the underlying crystal structure during van der Waals assembly, we first chose MoTe$_2$ as a growth template. MoTe$_2$ assumes various stable crystal phases, including 2H, 1T' and $T_d$ phases[7, 23] and we anticipate different assembly patterns of AgCN wires on hexagonal 2H and distorted 1T' phases of MoTe$_2$ crystals. Figure 1d-k confirms that the assembly behaviors of AgCN microwires grown on 2H- and 1T'-MoTe$_2$ are indeed different. The AgCN microwires grown on 2H-MoTe$_2$ (Figure 1d and 1i) display triaxial assembly, which is consistent with our previous observation[36]. On the other hand, we observed the uniaxial assembly of AgCN microwires on the distorted 1T'-MoTe$_2$ crystals as shown in Figure 1f and 1k. The detailed procedure to grow AgCN microwires on TMDC crystals can be found in the Methods section.

Although optical microscopic images showed the unidirectional alignment of AgCN on 1T'-MoTe$_2$, the azimuthal configuration of AgCN chains needs to be verified by crystal-direction-sensitive characterization methods. To achieve this goal, we performed polarized Raman spectroscopy (Figure 2), which has been widely utilized to identify orientations of distorted TMDCs[21-22, 24-25, 38]. Raman spectrum of MoTe$_2$ samples confirms the 1T' crystal phase[23-24] as shown in Supporting Information Figure S1. By careful analysis of polarized Raman spectra of 1T'-MoTe$_2$, we verified that the AgCN microwires are aligned along the distortion direction of the underlying 1T'-MoTe$_2$. We took a series of Raman spectra as a function of excitation laser polarization direction as shown in Figure 2c and 2d. In our data, $\theta = 0°$ corresponds to the case where the laser polarization is parallel to the AgCN microwires. Raman peaks at 78 cm$^{-1}$ and 163 cm$^{-1}$ indicate strong polarization dependence. In particular, the Raman peak at 163 cm$^{-1}$ location under 632.8 nm laser excitation showed the maximum intensity at $\theta = 0°$ (or $\theta = 180°$) configuration. Previous Raman studies have shown that the maximum intensity of 163 cm$^{-1}$ Raman peak of 1T'-MoTe$_2$ can be used to identify the distortion direction[39]. Since the observed distortion direction of 1T'-MoTe$_2$ from Raman is $\theta = 0°$ (AgCN microwire axis), we clearly confirmed that the AgCN microwire's axis is parallel to the distortion direction (b-axis) of the underlying 1T'-MoTe$_2$.

Electron diffraction and high-resolution TEM (HRTEM) imaging were also employed to directly visualize the relative orientation between the distorted MoTe$_2$ and AgCN microwire axis as shown in Figure 3. Figure 3a shows a TEM image of AgCN microwires grown on 1T'-MoTe$_2$.

Figure 3b shows the SAED pattern of AgCN microwires grown on 1T'-MoTe$_2$, where the 1T'-MoTe$_2$ crystal phase and AgCN microwire signals are labeled as red and blue circles, respectively. Two sets of diffraction signals were azimuthally aligned, which confirms that the atomic axis of AgCN microwires is parallel to the distortion direction (b-axis) of 1T'-MoTe$_2$. (Figure 3b). Therefore, the electron diffraction analysis is consistent with polarized Raman data presented in Figure 2.

HRTEM imaging analysis further established that the unidirectional alignment of AgCN chains is along the distorted lattice direction of underlying 1T'-MoTe$_2$. Figure 3d,g are zoomed-in STEM and TEM images displaying the lattice structures of 1T'-MoTe$_2$ and AgCN, respectively. Comparison with simulated STEM images (Figure 3e) verified that the vertical direction of the STEM image is parallel to the distortion direction (b-axis) of 1T'-MoTe$_2$. The atomic resolution TEM image of the AgCN lattice (Figure 3g and Supporting Figure S2) also shows the strong phase contrast pattern from the Ag position along the diagonal direction[36]. By comparing with the TEM simulation image (Figure 3h), we also confirmed that the AgCN atomic chain axis in the TEM image is aligned along the vertical direction.

The observed unidirectional alignment of AgCN microwires on 1T'-MoTe$_2$ can be generalized to similar phenomena on other distorted TMDC crystals. Figure 4 summarizes the behavior of the AgCN microwire assembly on T$_d$-WTe$_2$ and 1T'-ReS$_2$, which share the distorted crystal structure of 1T'-MoTe$_2$. Analysis with optical microscopy imaging, polarized Raman spectroscopy, TEM imaging, and SAED confirmed that the axis of AgCN microwires grown on WTe$_2$ and ReS$_2$ is also parallel to the distorted direction of underlying crystals. As shown in Figure 4a and 4f, AgCN microwires exhibit the unidirectional alignment on T$_d$-WTe$_2$ and 1T'-ReS$_2$ by optical microscope imaging. Polarized Raman spectroscopy of T$_d$-WTe$_2$ and 1T'-ReS$_2$ were also performed (Figure 4b,c and 4g,h). In particular, 165 cm$^{-1}$ of WTe$_2$ and 212 cm$^{-1}$ of ReS$_2$ exhibited the maximum Raman intensity along the laser polarization direction for the AgCN microwires ($\theta = 0°$). Previous polarized Raman studies have demonstrated that these Raman modes show maximum intensity along the distorted lattice direction[38, 40]. Through SAED and TEM analysis, the parallel configuration between the distorted lattice direction of WTe$_2$ (and ReS$_2$) and the AgCN microwire axis was also verified (Figure 4d,e and 4i,j).

By taking more than tens of SAED patterns from AgCN-grown various TMDC crystals, we prepared a histogram of azimuthal angles of AgCN microwires on distorted TMDCs as shown in Figure 5a. AgCN microwires exhibit exceptional rotational alignment along the distorted lattice directions for distorted octahedral TMDCs. The very narrow angular distribution (within 1° of error) indicates that the AgCN microwires can be an efficient way to identify the crystal orientation of distorted TMDCs. Previous works have indicated that ReS$_2$ exhibits rotational grains with 120°, 180°, or other tilt angles[19, 41]. We observed that the assembled AgCN microwires on exfoliated ReS$_2$ flakes occasionally show a domain pattern as shown in Figure 5b, where two local regions display different alignment directions of wire assembly. By performing polarized Raman spectroscopy, we confirmed that the observed change of AgCN assembly direction originates from the presence of rotational grains in ReS$_2$ crystals (Figure 5c). Different rotational grain structures of ReS$_2$ can be also identified, such as stripe-patterned alternating grains. Figure 5d shows that AgCN microwires display the alternating growth directions, which can be attributed to the presence of the stripe-patterned rotational grain structure. As shown in Figure 5e, the tentative assignment of grains 1 and 2 can be inferred from the observed epitaxy pattern of AgCN microwires. The assigned stripe grain pattern was confirmed with polarized Raman spectroscopy mapping (Figure 5f,g). The assembled AgCN wires can be easily removed by the treatment of ammonia solution as demonstrated in Figure S5 and a previous report[42]. Therefore, the assembly of AgCN microwires can be utilized as facile markers for the identification of crystal orientations of distorted TMDCs without the modification of TMDC's properties and surface quality.

It is worth noting that the observed unidirectional alignment behavior was not influenced by the lattice parameters of the underlying distorted 2D crystals (Table S1). Therefore, the observed general unidirectional alignment is likely due to the distortion-mediated van der Waals assembly of cyanide chains. Moreover, similar unidirectional alignment behavior was also observed with other kinds of cyanide chains (CuCN and $Cu_{0.5}Au_{0.5}CN$) on $T_d$-WTe$_2$, as shown in Figure S6. CuCN and $Cu_{0.5}Au_{0.5}CN$ share the same atomic chain structure with that of AgCN, but have slightly different lattice parameters along the chain direction[43]. Although the size and morphology of assembled wires depend on the cyanide chains, the unidirectional assembly behaviors of chains along the distortion of the substrate are maintained (Figure S6). Interestingly, the optical band gaps of metal cyanide chains (MCN) depend on the metallic species and can vary from approximately 2.3 eV (AuCN) to 4.5 eV (AgCN)[43]. The metal cyanides have potential as a component in the van der Waals heterostructures for various electrical and optoelectrical applications[35].

To understand the mechanism underlying the unidirectional assembly of AgCN wires on distorted TMDCs, we performed theoretical calculations for the adsorption energy of different AgCN chain configurations on ReS$_2$ (Figure 6a). In our calculations, the chain configuration with $N_1$ number of chains and $N_2$ number of AgCN units was labeled as $N_1 \times N_2$. Various chain configurations were generated on a sufficiently large ReS$_2$ monolayer substrate with a cut-off radius of ~10 Å. AgCN wires were shifted in a 40×40 xy-plane grid on the primitive 1T'-ReS$_2$ lattice and also translated in the z-direction at intervals of 0.01 Å around the reference vertical distance (5.0 Å from the Re atom). These parallel translations (x, y, and z directions) were repeated for every alignment azimuthal angle θ of AgCN wires on ReS$_2$, as shown in Figure 6a. The azimuthal angles were chosen at 15° intervals for computational efficiency. Finally, among various translated configurations, we identified the energetically most-favored configuration. The corresponding adsorption energy per AgCN unit is shown in Figure 6b.

From the azimuthal angle-dependent calculations, we clearly observed that the energetically favored configuration corresponds to $\theta = 0°$ for $N_2 \geq 3$ cases and the shape of the adsorption energy landscape is quite robust to different chain configurations. The side view of the most favored $1 \times 3$ configuration reveals that an AgCN chain prefers the hollow location between protruding S atom positions (Figure 6c). This is likely due to the larger degree of interaction between AgCN and S atoms at this position. Since the calculated favored energy configuration at $\theta = 0°$ mainly originates from the 1D lattice distortion, we expect a similar adsorption energy trend for 1T'-MoTe$_2$ and T$_d$-WTe$_2$. The calculated results for the adsorption energies of AgCN on 1T'-MoTe$_2$ also show a similar trend; the most favorable energy configuration occurs at $\theta = 0°$ (Figure S7). We note that the typical growth temperature ranges for AgCN assembly in our study are relatively low (< 100 °C), which may be one of the reasons for the observed remarkable degree of alignment of AgCN microwires[44].

**CONCLUSIONS**

In summary, we observed the unidirectional alignment of AgCN microwires on various distorted TMDC crystals. Through the combination of optical microscopy, polarized Raman spectroscopy, TEM analysis, and theoretical calculations, we demonstrated that metal cyanide chains are aligned along the distorted lattice direction (b-axis) of 1T'-MoTe$_2$, T$_d$-WTe$_2$, and 1T'-

ReS$_2$. The electrical properties of MoTe$_2$ are strongly influenced by the crystal phases; thus, a simple method to distinguish different crystal phases is valuable. Our observations clearly indicate that the surface-mediated van der Waals assembly of AgCN microwires can effectively distinguish between 2H and 1T' phases of MoTe$_2$. A simple observation of AgCN wire assembly patterns via optical microscopy can serve as an effective and facile way to differentiate various phases and orientations of the TMDC family. We envision that the observed unidirectional assembly phenomena on distorted TMDCs can be generalized to other assembly components including metallic films[45-46] and hybrid perovskites[44].

**METHODS**

**Sample preparation.** 1T'-MoTe$_2$, T$_d$-WTe$_2$ (Supporting Figure S3), and 1T'-ReS$_2$ (Supporting Figure S4) were purchased from HQ Graphene. 1T'-MoTe$_2$ (WTe$_2$, ReS$_2$) crystals were prepared through mechanical exfoliation on 280 nm-thick SiO$_2$/Si substrates. AgCN microwires were grown on exfoliated 2D crystals using the previously developed drop-cast method[35-36]. In brief, a solution of AgCN (1~2 mM) was prepared by dissolving AgCN (99%, Sigma-Aldrich) in ammonia solution (14.8 M, Samchun). Using a micro-pipet, 1.5μL of AgCN solution was dropped onto target TMDC crystals on SiO$_2$/Si wafers. Mild heating on a hot-plate at 85℃ was performed for 15 minutes. The assembly of CuCN (0.6 mM) or Au$_{0.5}$Cu$_{0.5}$CN (0.6 mM) wires was performed using the same growth method. To remove AgCN microwires, the samples were placed in ammonia solution for five minutes and rinsed in DI water. TEM samples were prepared by transferring AgCN-grown 2D crystals by direct transfer method, mediated by isopropyl alcohol drying[36].

**Optical imaging and electron microscopy.** Optical microscope images were acquired using a Leica DM-750M with visible light. TEM images and SAED patterns were obtained with Cs-corrected JEOL ARM-200F operated at 200kV. TEM image simulation was performed using MacTempasX Version 2 software. TEM imaging acquisition conditions including a Cs value of −12 μm, convergence angle value of 0.1 mrad, mechanical vibration value of 0.5 Å, and defocus values (−2 nm for AgCN) were used for simulations of AgCN. STEM images of 1T'-$MoTe_2$ were simulated using inner and outer apertures of 90 mrad and 370 mrad, respectively.

**Polarized Raman spectroscopy.** The Raman measurements were performed by using a home-built confocal micro-Raman system using the 1.96 eV (632.8 nm) line of a He-Ne laser. A 50× objective lens (0.8 NA) was used to focus the laser beam onto the sample and to collect the scattered light (backscattering geometry). The Raman signal was dispersed with a HORIBA iHR550 spectrometer with a grating of 2400 grooves per millimeter and detected and a liquid-nitrogen-cooled back-illuminated charge-coupled-device (CCD) detector. The laser power was kept below 0.1 mW to avoid heating. Three volume holographic notch filters (OptiGrate) were used to observe the low frequency region (<100 $cm^{−1}$). An achromatic half-wave plate was used to rotate the polarization of the linearly polarized laser beam to the desired direction. All measurements were conducted in the parallel-polarization configuration, where the analyzer angle was set such that photons with the polarization parallel to the incident polarization pass through. Another achromatic half-wave plate was placed in front of the spectrometer to keep the polarization direction of the signal entering the spectrometer constant with respect to the groove direction of the grating.

**First-principle calculations.** We calculated the potential energy surface through the L-J potential[36] to determine the optimal growth direction of AgCN wires on 1T'-ReS$_2$, 1T'-MoTe$_2$, and 2H-MoTe$_2$ substrates. We used exhaustive search with UFF potential[47]. The lattice parameters of AgCN, ReS$_2$, 1T'-MoTe$_2$, 2H-MoTe$_2$ and the reference vertical distance between AgCN and substrates were obtained by density functional theory (DFT)[48] calculations as implemented in the Vienna Ab-initio Simulation Package (VASP)[49] with the projector augmented wave (PAW) method[50]. PBE exchange-correlation functional[51] based on the generalized gradient approximation (GGA) was employed. Considering the weak interactions between AgCN wires and substrates, zero-damping DFT-D3 correction[52] was used. 600 eV of cut-off energy and 8×8×8 K-points generated in the Monkhorst-Pack scheme[53] were used to optimize the bulk lattice parameters. The optimizations were carried out until the Hellmann-Feynman force acting on each atom was less than 0.01 eV/Å. For estimation of interlayer distance, AgCN monomer and dimer were placed on the sufficiently large ReS$_2$ (or 1T'-MoTe$_2$) monolayer with ~10 Å of spacing between neighboring AgCN in a supercell scheme, with ~10 Å of vacuum slab. These calculations were performed with 400 eV of cut-off energy and only Γ-point in reciprocal space.

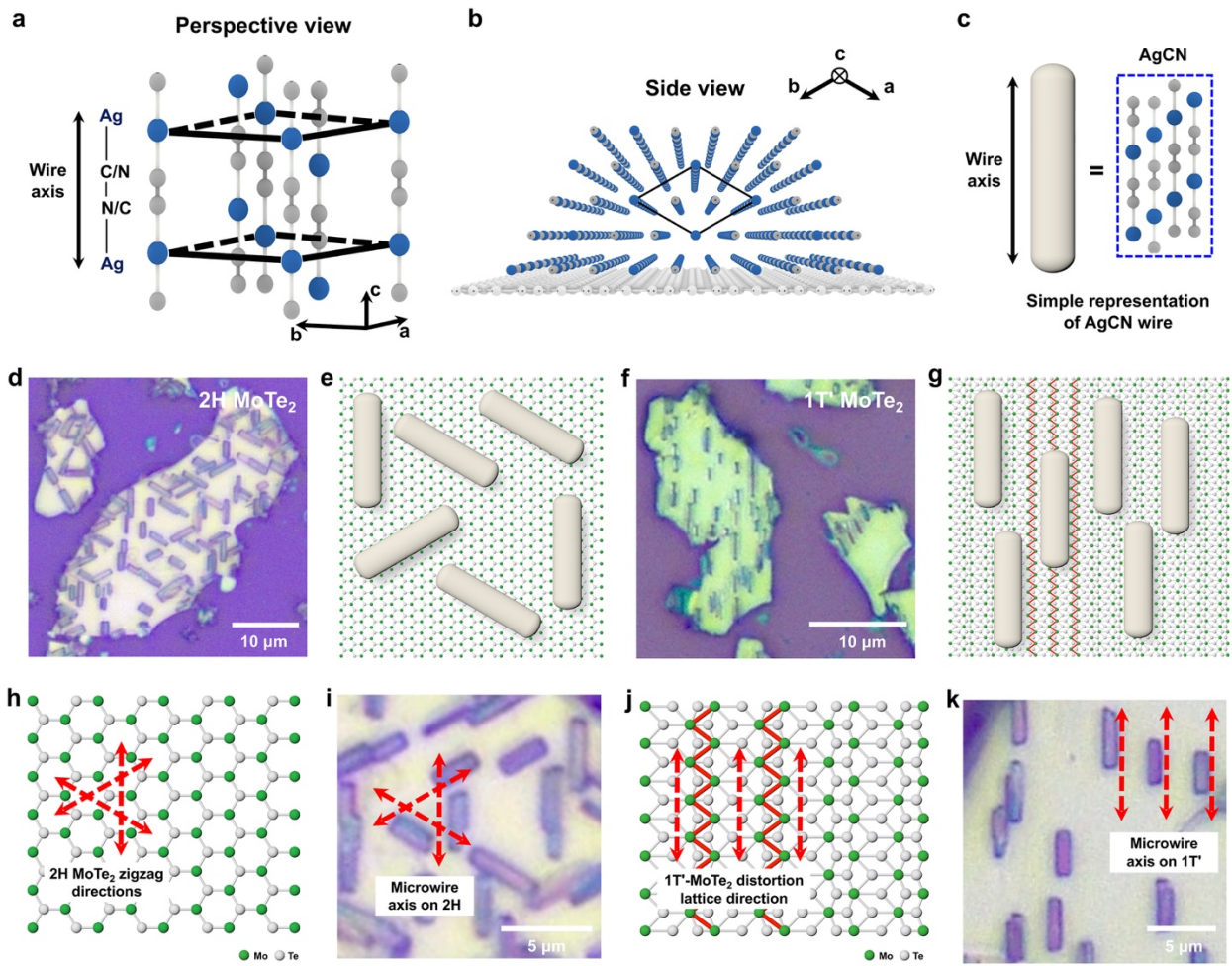

**Figure 1. AgCN chain assembly on 2H- and 1T'-MoTe$_2$.** a,b) Perspective and cross-sectional atomic model of AgCN assembly. c) Simplified representation of AgCN chain assembly as a microwire. d) Optical image of AgCN microwires grown on 2H-MoTe$_2$. e) Schematic diagram of AgCN wire grown on 2H-MoTe$_2$. f) Optical image of AgCN microwires on 1T'-MoTe$_2$. g) Schematic diagram of AgCN wire grown on 1T'-MoTe$_2$. h,i) Atomic model of 2H-MoTe$_2$ and zoomed-in optical image of AgCN wires on 2H-MoTe$_2$. j,k) Atomic model of 1T'-MoTe$_2$ and zoomed-in optical image of AgCN wires on 1T'-MoTe$_2$.

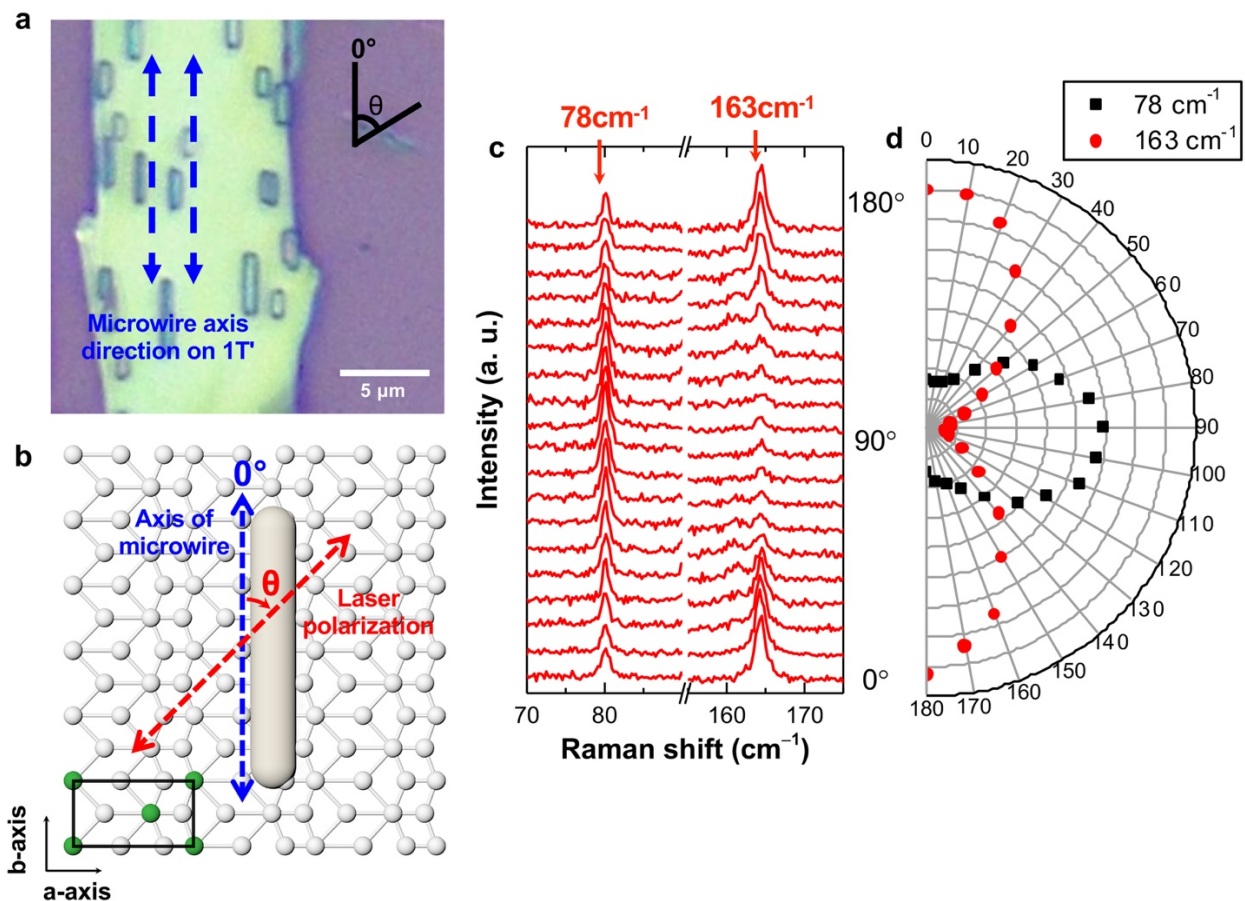

**Figure 2. Polarized Raman spectroscopy of 1T'-MoTe$_2$ for identification of azimuthal alignment angle of AgCN wire axis.** a) Optical image of 1T'-MoTe$_2$ crystal with AgCN microwires used for polarized Raman spectroscopy. b) Schematic illustration of Raman spectroscopy with polarized excitation. Polarization direction $\theta$ is set as a relative angle from the observed AgCN microwire axis. c) Polarization angle dependent Raman spectra of 1T'-MoTe$_2$. d) Polar plot of Raman peak intensity of 1T'-MoTe$_2$.

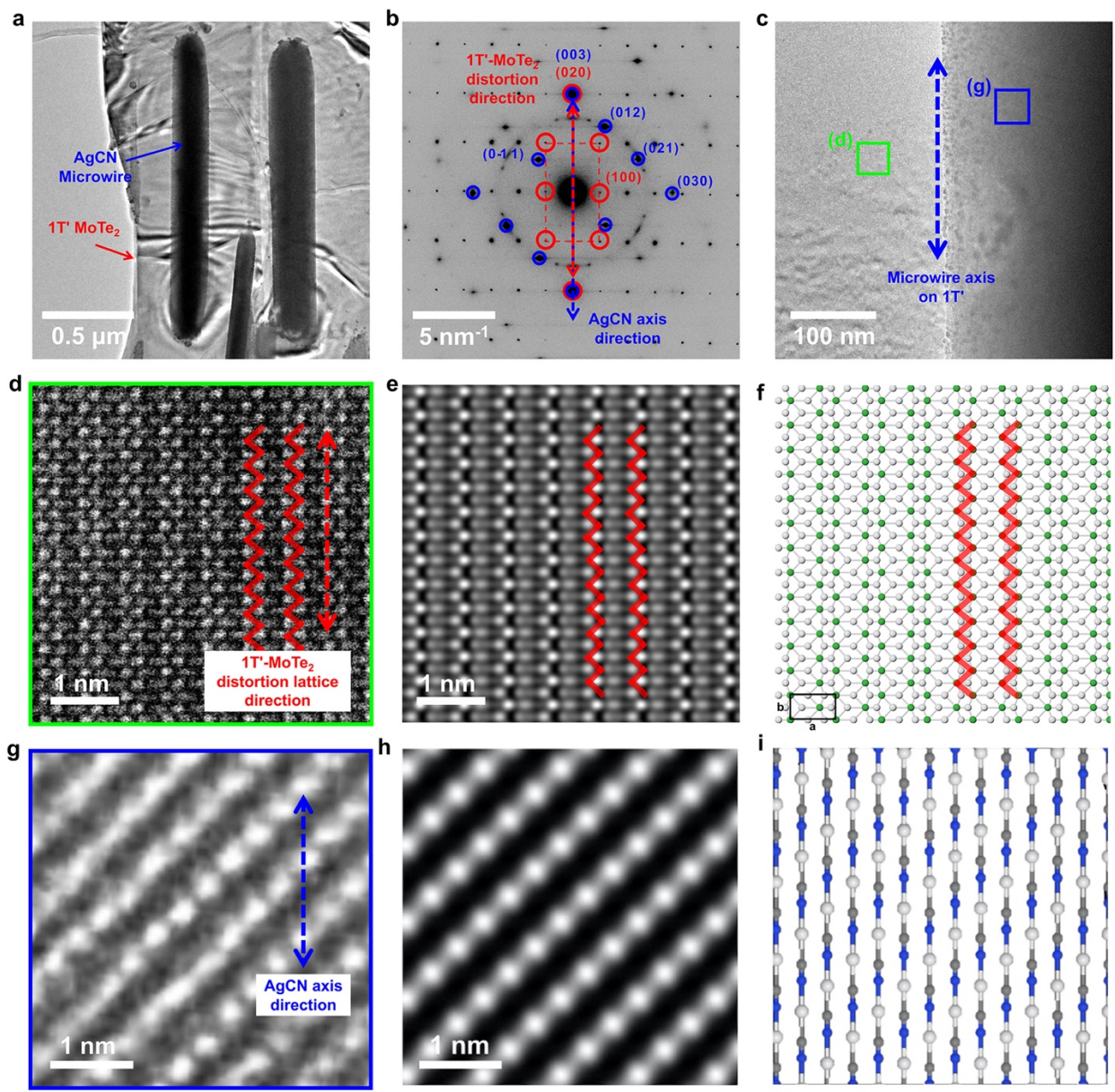

**Figure 3. TEM imaging and electron diffraction of AgCN microwires grown on 1T'-MoTe$_2$.** a) TEM image of AgCN microwire grown on 1T'-MoTe$_2$. b) SAED of AgCN wires on 1T'-MoTe$_2$. c) TEM image of a AgCN wire on 1T'-MoTe$_2$. Green and blue boxes are fields of view for panel d and g. d) Zoomed-in STEM image of 1T'-MoTe$_2$. e) STEM simulation image of 1T'-MoTe$_2$. f) Atomic model of 1T'-MoTe$_2$. g) Zoomed-in TEM image of AgCN microwire. h) Simulation image of AgCN. i) Atomic model of AgCN used for simulation.

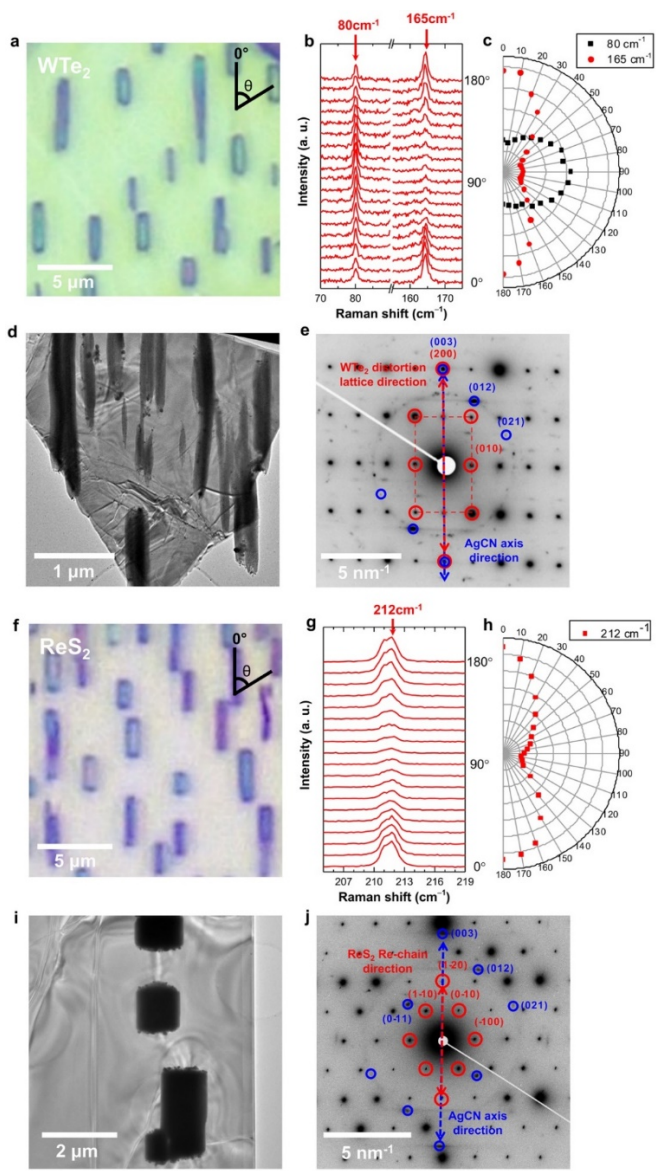

**Figure 4. Unidirectional alignment of AgCN microwires on T$_d$-WTe$_2$ and 1T'-ReS$_2$.** a) Optical image of AgCN microwires grown on WTe$_2$. b) Polarization angle dependent Raman spectra of WTe$_2$. c) Polar plot of Raman peak intensity of WTe$_2$. d) TEM image of AgCN microwires grown on WTe$_2$. e) Electron diffraction of AgCN microwires grown on WTe$_2$. f) Optical image of AgCN microwires grown on ReS$_2$. g) Polarization angle dependent Raman spectra of ReS$_2$. h) Polar plot of Raman peak intensity of ReS$_2$. i) TEM image of AgCN microwires grown on ReS$_2$. j) Electron diffraction of AgCN microwires grown on ReS$_2$.

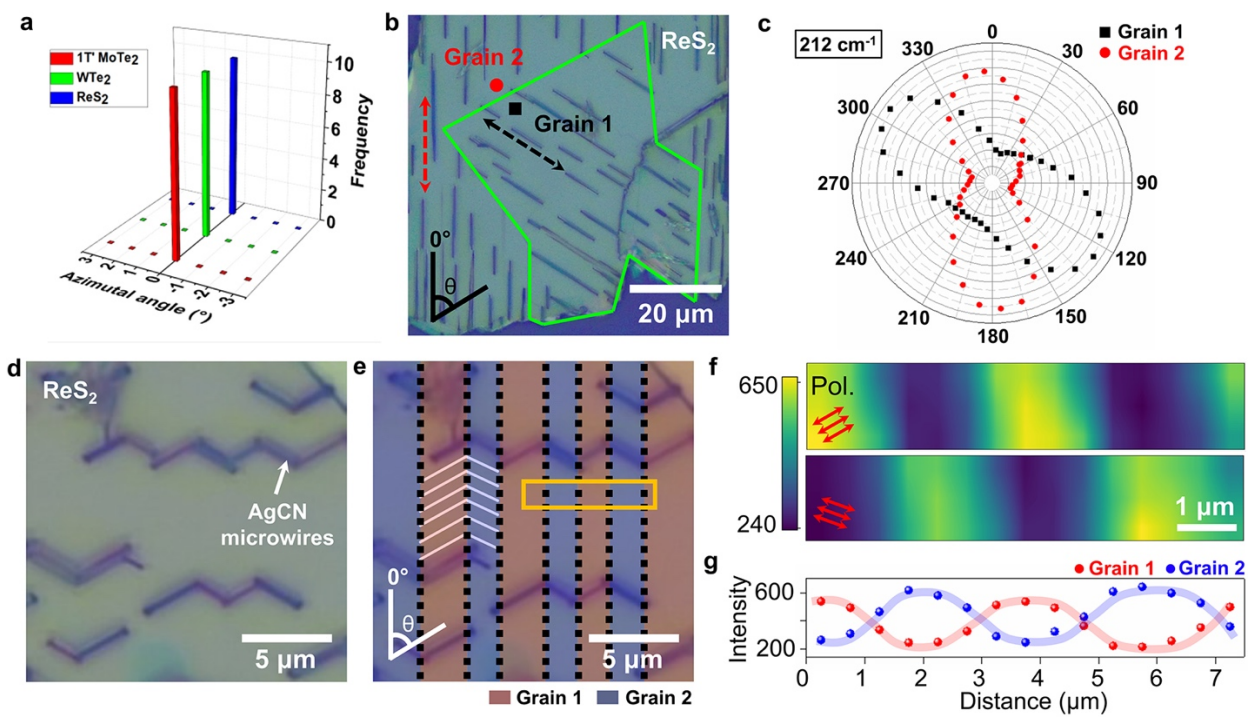

**Figure 5. AgCN assembly for visualization of ReS₂ rotational grains.** a) Histogram of measured azimuthal alignment angles $\theta$ of AgCN chains on distorted TMDCs. b) Optical image of AgCN microwires grown with two distinct directions on ReS$_2$. c) Polar plot of Raman peak (212 cm$^{-1}$) intensity of ReS$_2$ at marked two locations in panel b. d) Optical image of AgCN microwires grown with alternating two directions on ReS$_2$. e) Assignment of ReS$_2$ grains from the observed AgCN assembly pattern. The rectangle is the field of view for polarized Raman mapping in panels f. f) Polarized Raman peak intensity mapping at polarization directions of 60° (top) and 110° (bottom). g) Line profiles of polarized Raman intensities in panel f along the horizontal direction.

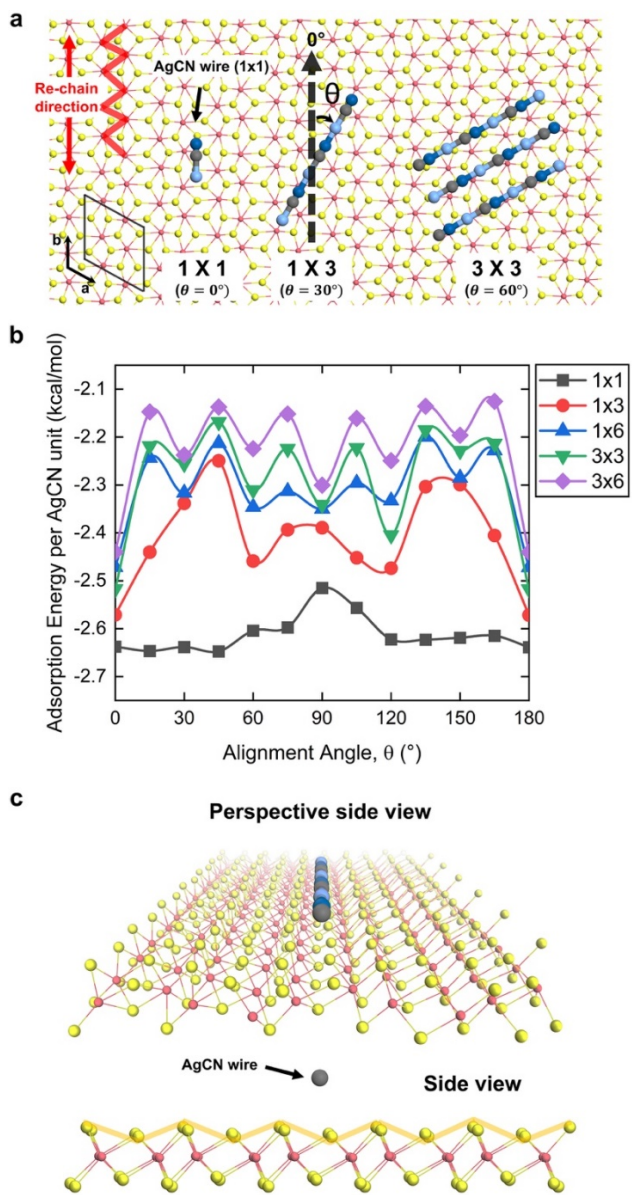

**Figure 6. Theoretical calculations on unidirectional assembly behavior of AgCN chains on distorted crystals.** a) AgCN chain configurations on ReS$_2$. The number of Ag−C≡N unit and azimuthal alignment angle were varied for calculations. b) Adsorption energy of AgCN chains on ReS$_2$ as a function of azimuthal alignment angle $\theta$. c) Energetically favored AgCN wire (1 × 3) geometry positioned on the distorted lattice of ReS$_2$.

ASSOCIATED CONTENT

**Supporting Information**.

The Supporting Information is available free of charge on the ACS Publications website.

Extra experimental Raman spectroscopy data, high resolution STEM image of MoTe$_2$, removal of AgCN microwires, assembly of CuCN and Cu$_{0.5}$Au$_{0.5}$CN on WTe$_2$, theoretical calculations for AgCN on 1T'- and 2H-MoTe$_2$, and summary on the lattice parameters of 2D crystals in this work.

AUTHOR INFORMATION

**Corresponding Author**

Email: kpkim@yonsei.ac.kr

**Author Contributions**

M.J. and K.K. conceived the project and designed the experiments. M.J. mainly performed the experiments and analyzed the data. H.B. and H.L. performed first principles calculations. W.N. and H.C. performed polarized Raman measurements. Y.L. and B.Y. performed TEM and STEM imaging. S.C. performed image simulation of TEM and STEM. All authors discussed the results and commented on the manuscript.

**Notes**

The authors declare no competing financial interest.

ACKNOWLEDGMENT

This work was mainly supported by the Basic Science Research Program through the National Research Foundation of Korea (NRF-2017R1A5A1014862, 2019R1A2C3006189 and NRF-2019R1C1C1003643) and by the Institute for Basic Science (IBS-R026-D1). Y.L. received support from the Basic Science Research Program at the National Research Foundation of Korea which was funded by the Ministry of Education (NRF-2020R1A6A3A13060549) and from the 2020 Yonsei University Graduate School Research Scholarship Grants.